\documentstyle[amssymb,preprint,aps]{revtex}

\begin{document}
\title{Evolutionarily Stable Strategies in Quantum Games}
\author{A. Iqbal and A.H.Toor}
\address{Department of Electronics, Quaid-e-Azam University, Islamabad, Pakistan.\\
email: qubit@isb.paknet.com.pk}
\date{\today }
\maketitle

\begin{abstract}
Evolutionarily Stable Strategy (ESS) in classical game theory is a
refinement of Nash equilibrium concept. We investigate the consequences when
a small group of mutants using quantum strategies try to invade a classical
ESS in a population engaged in symmetric bimatrix game of Prisoner's
Dilemma. Secondly we show that in an asymmetric quantum game between two
players an ESS pair can be made to appear or disappear by resorting to
entangled or unentangled initial states used to play the game even when the
strategy pair remains a Nash equilibrium in both forms of the game.
\end{abstract}

\section{Introduction}

An evolutionarily stable strategy (ESS) was originally defined by Maynard
smith and price (1973) [1] with the motivation that a population playing the
ESS can withstand a small, invading group. This concept was developed by
combining ingredients from game theory and from some work on the evolution
of the sex ratio. Maynard Smith considers a population in which members are
matched randomly in pairs to play a bimatrix game. The players are
anonymous, that is any pair of players plays the same symmetric bimatrix
game and the players are identical with respect to their set of strategies
and their payoff function. From the symmetry of the bimatrix game it is
meant that for strategy set $S$ the payoff of the first player when he plays 
$A\in S$ and his opponent plays $B\in S$ is same as the payoff of the second
player when the latter plays $A$ and the first player plays $B$. A
population which adopts an ESS can withstand a small invading group [1--3].
An evolutionarily game means a model of strategic interaction continuing
over time in which higher payoff strategies gradually displace strategies
with lower payoffs. There is also some inertia involved in distinguishing
between evolutionarily and revolutionary changes; here inertia means that
aggregate behavior does not change too abruptly.

In a recent publication, Meyer [4] examined game theory from a quantum
perspective and showed that a player can enhance his expected payoff by
implementing a quantum strategy. Shortly afterwards, Eisert et al [5]
investigated the quantization of nonzero sum game, the Prisoner's Dilemma
(PD), and showed that, if quantum strategies are admitted, the dilemma no
longer exists; moreover, they succeeded in constructing a particular quantum
strategy which will always outmaneuver any classical strategy. Referring to
Dawkins' `Selfish Gene'[6], these authors also hinted that games of survival
are being played already on the molecular level, where quantum mechanics
dictates the rules. Coming back to the concept of an ESS we notice that it
was perhaps introduced into classical game theory for two reasons:

1.\qquad Two player games can have multiple Nash equilibria (NE).

2.\qquad Population biology problems can be modelled with the help of this
concept.

The reasons for claim that (1) holds for quantum as well as classical games
are not far from obvious. In our opinion even the reason (2) may have a
meaning in quantum setting. In section $(ii)$ of this paper it is our
purpose to take up the idea that `games of survival are played at molecular
level' and see what happens when `mutants' of ESS theory come up with
quantum strategies and try to invade classical ESS. What happens if such an
invasion is successful and a new ESS is established; an ESS that is quantum
in nature? Suppose afterwards another small group of `mutants' appear
equipped with some other quantum strategy. Would it be successful now to
invade the quantum ESS? These questions have been considered in section $%
(ii) $ for the pairwise symmetric game of PD.

We are trying to extend an idea originally proposed for problems in
population biology to quantum domain and it needs more substantive evidence
than we have provided. Our motivation is to look for the consequences when
quantum strategies that Eisert et al have called one and two-parameter
quantum strategies satisfy the criteria of an ESS. Later we also consider NE
and ESS's in the PD game when it is played via another scheme proposed
recently by Marinatto and Weber [13].

PD is a symmetric game and the payoff to a player depends only on player's
strategy not on player's identity. In section $(iii)$ we consider
quantization of a general $2\times 2$ matrix game in asymmetric form played
via Marinatto and Weber's scheme. In evolutionary game theory an ESS for
such a game is a strategy pair that forms a strict Nash equilibrium (NE)
[11].We also search what should exactly be the initial entangled state to
play a quantum game when a particular strategy pair is a NE in both the
classical and quantum versions of the game but an ESS in only version.

\section{Symmetric case}

The PD game has classical available pure strategies Cooperation ($C$) and
Defection ($D$) [7]. An interesting question is which strategies are likely
to be stable and persistent in a population engaged in the pairwise version
of the game. A simple analysis [8] show that $D$ will be the pure classical
strategy prevalent in the population and hence the classical ESS. In
general, suppose that a strategy $A$ is played by almost all members of the
population, the rest of the population form a small group of mutants playing
strategy $B$ constitute a fraction $\epsilon $ of the total population. The
strategy $A$ is said to be evolutionarily stable (ES) [9] against $B$ if $%
P[A,(1-\epsilon )A+\epsilon B]>P[B,(1-\epsilon )A+\epsilon B]$ where $P[A,B]$
is defined as the payoff to player playing $A$ against player playing $B$,
for all sufficiently small, positive $\epsilon $.There exists some $\epsilon
_{0}$, such that for $\epsilon \in \lbrack 0,\epsilon _{0})$ the inequality
is satisfied [9]. If for the given $A$ and $B$ the $\epsilon _{0}$ specified
is as large as possible the $\epsilon _{0}$ is called the ``invasion
barrier''. If $B$ comes at a frequency larger than $\epsilon _{0}$ it will
lead to an invasion.

For a symmetric bimatrix game it follows [8,9] that $A$ is an ESS with
respect to $B$

1.\qquad If $P[A,A]>P[B,A]\qquad $and

2.\qquad If $P[A,A]=P[B,A]$ then $P[A,B]>P[B,B]\qquad \qquad (1)$

If most of the players play $A$, then almost all potential opponents are $A$
players, so if $A$ does better against $A$ than $B$ does, $B$ players will
be persistent losers as the game evolves. However, if $A$ and $B$ do equally
well against $A$, then how well the strategies perform against $B$ becomes
important. Therefore, for $A$ to be ES against $B$ the strategy $A$ must
then do better against $B$ than $B$ does against $B$. Strategy $A$ is an ESS
if $A$ is ES against all $B\neq A$. For pure strategies $A$ and $B$
(classical as well as quantum) the fitnesses [8] can be defined as

$W(A)=P(A,A)F_{A}+P(A,B)F_{B}\qquad W(B)=P(B,A)F_{A}+P(B,B)F_{B}\qquad
\qquad (2)$

Where $F_{A}$ and $F_{B}$\ are the classical frequencies of the pure
strategies $A$ and $B$ respectively. A quantum strategy cannot be treated as
a probabilistic sum of pure classical strategies (except under special
conditions). Therefore for finding fitness the quantum strategies are
treated as `new' strategies that cannot be reduced to the pure classical
strategies.

An ESS is usually considered another refinement of the NE concept. For
symmetric bimatrix games the relationship is described as [11]

$\bigtriangleup ^{ESS}\subset \bigtriangleup ^{PE}\subset $ $\bigtriangleup
^{NE}$ and $\bigtriangleup ^{PE}\neq \Phi $ where $\bigtriangleup
^{NE},\bigtriangleup ^{PE}$and $\bigtriangleup ^{ESS}$are the sets of
symmetric NE, symmetric proper equilibrium and evolutionarily stable
strategies respectively. Application of quantum theory gives a new set of NE
strategies $\bigtriangleup ^{NE}$and $\bigtriangleup ^{ESS}$may contain
quantum strategies as well.

We assume the same quantum version of PD game as described by Eisert et al
[5] between two players. A pair of qubits are prepared in unentangled state $%
\left| CC\right\rangle $ and sent through the entangling gate $\widehat{J}$. 
$\widehat{J}$ is essentially a unitary operator known to both players and is
symmetric with respect to the interchange of two players. The two players,
call them Alice and Bob, then apply their local unitary operators $U_{{\rm A}%
}$ and $U_{{\rm B}}$ respectively. An inverse gate to $\widehat{J}$ is
applied before the final measurement by the arbiter. Let $s_{{\rm A}}$ and $%
s_{{\rm B}}$ be Alice's \ and Bob's strategies respectively. The payoff
matrix is the same as chosen by Eisert et al [5] and can be written as

$\left( 
\begin{array}{cc}
(3,3) & (0,5) \\ 
(5,0) & (1,1)
\end{array}
\right) \qquad \qquad (3)$

Suppose the players apply their respective strategies $s_{{\rm A}}$ and $s_{%
{\rm B}}$. These strategies are unitary operators at player's disposal i.e. $%
s_{{\rm A}}\symbol{126}U_{{\rm A}}$ and $s_{{\rm A}}\symbol{126}U_{{\rm B}}.$%
If initial state is maximally entangled state $\rho $ then the final
state[4] is

$\sigma =(U_{{\rm A}}\otimes U_{{\rm B}})\rho (U_{{\rm A}}\otimes U_{{\rm B}%
})^{\dagger }\qquad \qquad (4)$

The arbiter applies Kraus operators [4,10] on $\sigma $

$\pi _{CC}=\left| \psi _{CC}\right\rangle \left\langle \psi _{CC}\right|
\qquad \pi _{CD}=\left| \psi _{CD}\right\rangle \left\langle \psi
_{CD}\right| $

$\pi _{DC}=\left| \psi _{DC}\right\rangle \left\langle \psi _{DC}\right|
\qquad \pi _{DD}=\left| \psi _{DD}\right\rangle \left\langle \psi
_{DD}\right| \qquad \qquad (5)$

The expected payoffs to Alice and Bob are [4]

$P_{A,B}=[P_{CC}]_{A,B}tr[\pi _{CC}\sigma ]+[P_{CD}]_{A,B}tr[\pi _{CD}\sigma
]$

$\qquad +[P_{DC}]_{A,B}tr[\pi _{DC}\sigma ]+[P_{DD}]_{A,B}tr[\pi _{DD}\sigma
]\qquad \qquad (6)$

Because the game is symmetric we define $P(C,D)$ as the payoff to $C$ player
against $D$ player. Similarly $P(D,C)$ is defined. The subscripts of A \ and
B are not required.

Eisert et al [4] have used following matrix representations of the unitary
operators of one and two-parameter strategies respectively.

$U(\theta )$=$\left( 
\begin{array}{cc}
\cos \text{(}\theta \text{/2)} & \sin \text{(}\theta \text{/2)} \\ 
\text{-}\sin \text{(}\theta \text{/2)} & \cos \text{(}\theta \text{/2)}
\end{array}
\right) \qquad U(\theta ,\phi )=\left( 
\begin{tabular}{ll}
e$^{i\phi }\cos \text{(}\theta \text{/2)}$ & $\sin \text{(}\theta \text{/2)}$
\\ 
$\text{-}\sin \text{(}\theta \text{/2)}$ & e$^{-i\phi }\cos \text{(}\theta 
\text{/2)}$%
\end{tabular}
\right) \qquad (7)$

Where $\theta \in \lbrack 0,\pi ]$ and $\phi \in \lbrack 0,\pi /2]$. The
classical pure strategies $C$ and $D$ are realized as $C\symbol{126}U(0)$, $D%
\symbol{126}U(\pi )$ respectively for one-parameter strategies and $C\symbol{%
126}U(0,0)$, $D\symbol{126}U(\pi ,0)$ respectively for two-parameter
strategies. We consider three cases:

A.\qquad A small group of mutants appear equipped with one-parameter quantum
strategy $U(\theta )$ when $D$ exists as a classical ESS.

B.\qquad The mutants are equipped with two-parameter quantum strategy $%
U(\theta ,\phi )$ against classical ESS.

C.\qquad The mutants have successfully invaded and a two-parameter quantum
strategy $Q\symbol{126}U(0,\pi /2)$ has established itself as a new quantum
ESS. Again another small group of mutants appear using some other
two-parameter quantum strategy and try to invade the quantum ESS $Q$.

{\bf Case A}

The expected payoffs are found as

$P(\theta ,D)=\sin ^{2}(\theta /2)$

$P(\theta ,\theta )=2\cos ^{2}(\theta /2)+5\cos ^{2}(\theta /2)\sin
^{2}(\theta /2)+1$

$P(D,\theta )=5\cos ^{2}(\theta /2)+\sin ^{2}(\theta /2)$

$P(D,D)=1\qquad \qquad (8)$

Where, for example $P(\theta ,D)$ is the payoff to mutant employing
one-parameter strategy $\theta $ against the opponent using $D$. Now $%
P(D,D)>P(\theta ,D)$ for all $\theta \in \lbrack 0,\pi )$. Hence, the first
condition for an ESS is satisfied and $D\symbol{126}U(\pi )$ is an ESS. The
case $\theta =\pi $ corresponds to the case when one-parameter mutant
strategy coincides with the ESS and is ruled out. If $D\symbol{126}U(\pi )$
is played by almost all the members of the population, which correspond to
high frequency $F_{D}$ for $D$, we have then $W(D)>W(\theta )$ for all $%
\theta \in \lbrack 0,\pi )$. Therefore the fitness of a one-parameter
quantum strategy, which also corresponds to the case of mixed (randomized)
classical strategies [4], cannot be greater than that of a classical ESS. A
one-parameter quantum strategy, therefore, cannot succeed to invade a
classical ESS.

{\bf Case B}

The expected payoffs are

$P(D,D)=1$

$P(D,U)=5\cos^{2}(\phi )\cos ^{2}(\theta /2)+\sin ^{2}(\theta /2)$

$P(U,D)=5\sin^{2}(\phi )\cos ^{2}(\theta /2)+\sin ^{2}(\theta /2)$

$P(U,U)=3\left| \cos (2\phi )\cos ^{2}(\theta /2)\right| ^{2}+5\cos
^{2}(\theta /2)\sin ^{2}(\theta /2)\left| \sin (\phi )-\cos (\phi )\right|
^{2}$

$\qquad \qquad +\left| \sin (2\phi )\cos ^{2}(\theta /2)+\sin ^{2}(\theta
/2)\right| ^{2}\qquad \qquad (9)$

Here $P(D,D)>P(U,D)$ if $\phi <\arcsin (1/\sqrt{5})$ and if $P(D,D)=P(U,D)$
then $P(D,U)>P(U,U)$. Therefore $D$ is an ESS if $\phi <\arcsin (1/\sqrt{5})$
otherwise the strategy $U(\theta ,\phi )$ will be in position to invade $D$.
Alternatively if most of the members of the population play $D\symbol{126}%
U(\pi ,0)$, meaning high frequency $F_{D}$ for $D$, then the fitness $W(D)$
will remain greater than the fitness $W[U(\theta ,\phi )]$ if $\phi <\arcsin
(1/\sqrt{5})$. For $\phi >\arcsin (1/\sqrt{5})$ the strategy $U(\theta ,\phi
)$ can invade the strategy $D$ which is an ESS. The possession of a richer
strategy by the mutants in this case leads to an invasion of $D$ when $\phi
>\arcsin (1/\sqrt{5})$. Mutants having access to richer strategies may seem
non-judicious but \ even in classical setting an advantage by the mutants
leading to invasion may be seen in similar context.

{\bf Case C}

Eisert et al [4] showed that the quantum strategy $Q\symbol{126}U(0,\pi /2)$
played by both the players is the unique NE and one player cannot gain
without lessening the other player's expected payoff. The expected payoffs
are

$P(Q,Q)=3$

$P(U,Q)=[3-2cos^{2}(\phi )]\cos ^{2}(\theta /2)$

$P(Q,U)=[3-2cos^{2}(\phi )]\cos ^{2}(\theta /2)+5\sin ^{2}(\theta /2)\qquad
\qquad (10)$

Now $P(Q,Q)>P(U,Q)$ holds true for all $\theta \in \lbrack 0,\pi ]$ and $%
\phi \in \lbrack 0,\pi /2]$ except when $\theta =0$ and $\phi =\pi /2$ which
is the case when the mutant strategy $U(\theta ,\phi )$ is the same as $Q$
and is ruled out. Therefore the first condition for $Q$ to be an ESS is
satisfied. The condition $P(Q,Q)=P(U,Q)$ implies $\theta =0$ and $\phi =\pi
/2$. We have again the situation of the mutant strategy to be same as $Q$
and we neglect it. If $Q$ is played by most of the players, meaning high
frequency $F_{Q}$ \ for $Q$, then it is seen that $W(Q)>W[U(\theta ,\phi )]$
\ for all $\theta \in (0,\pi ]$ and $\phi \in \lbrack 0,\pi /2)$. Therefore
a two parameter quantum strategy $U(\theta ,\phi )$ cannot invade the
quantum ESS i.e. the strategy $Q\symbol{126}U(0,\pi /2)$ for this particular
game. The mutants having access to richer strategy space remains an
advantage not any more now. For the population as well as the mutants $Q$ is
the unique NE and ESS of the game.

The invasion of the mutants in case B does not seem so unusual given the
richer structure of strategy space they exploit and they are unable to
invade when it doesn't remain an advantage and most of the population have
access to it.

We now see what happens to PD game when played via Marinatto's scheme[13].
In this scheme the players apply their `tactics' by restricting themselves
to a probabilistic choice between the identity operator $\stackrel{%
\curlywedge }{I}$ and the Pauli spin-flip operator $\stackrel{\curlywedge }{%
\sigma _{x}\text{.}}$The purpose [12] for such a choice as described by the
authors is to have the smallest set of operations able to reproduce, when
applied to a factorizeable couple of strategies, the results of the
classical theory of games. However new results come out from the richer
structure of the strategic space, i.e. from the entangled couple of
strategies[13]. S.C.Benjamin in his comment [14] have considered it a severe
restriction on the full range of quantum mechanically possible manipulations
but Marinatto and Weber have replied [12] by describing it a `minimal'
choice enough to reproduce the classical results.

For the initial entangled state

$\left| \psi _{in}\right\rangle =a\left| CC\right\rangle +b\left|
DD\right\rangle \qquad \qquad \left| a\right| ^{2}+\left| b\right|
^{2}=1\qquad \qquad (11)$

when $\stackrel{\curlywedge }{I}$ and $\stackrel{\curlywedge }{\sigma _{x}}$
correspond to strategies $C$ and $D$ respectively with the payoff matrix
(3). Payoffs to Alice and Bob are:

$P_{A}(p,q)=3\{pq\left| a\right| ^{2}+(1-p)(1-q)\left| b\right|
^{2}\}+5\{p(1-q)\left| b\right| ^{2}+q(1-p)\left| a\right| ^{2}\}$

$\qquad \qquad +\{pq\left| b\right| ^{2}+(1-p)(1-q)\left| a\right| ^{2}\}$

$P_{B}(p,q)=3\{pq\left| a\right| ^{2}+(1-p)(1-q)\left| b\right|
^{2}\}+5\{p(1-q)\left| a\right| ^{2}+q(1-p)\left| b\right| ^{2}\}$

$\qquad \qquad +\{pq\left| b\right| ^{2}+(1-p)(1-q)\left| a\right|
^{2}\}\qquad \qquad (12)$

where $p$ and $q$ are the probabilities of Alice and Bob respectively to act
with the operator $\stackrel{\curlywedge }{I}$. PD is symmetric game and
remains symmetric after quantizing it. For a symmetric bimatrix games an ESS
is recognized as a symmetric NE with an additional property usually called
`the stability property'[8].

We search for symmetric NE from the inequalities using only the parameter $b$
of the initial state $\left| \psi _{in}\right\rangle $ because for the state 
$\left| \psi _{in}\right\rangle =a\left| CC\right\rangle +b\left|
DD\right\rangle $ the game reduces to classical when $\left| b\right| ^{2}=0$
i.e. when the initial state becomes unentangled. NE inequalities are then

$P_{A}(\stackrel{\star }{p},\stackrel{\star }{q})-P_{A}(p,\stackrel{\star }{q%
})=(\stackrel{\star }{p}-p)\{3\left| b\right| ^{2}-(\stackrel{\star }{q}%
+1)\}\geq 0$

$P_{B}(\stackrel{\star }{p},\stackrel{\star }{q})-P_{B}(\stackrel{\star }{p}%
,q)=(\stackrel{\star }{q}-q)\{3\left| b\right| ^{2}-(\stackrel{\star }{p}%
+1)\}\geq 0\qquad \qquad (13)$

The parameters of the initial entangled state $a$ and $b$ may decide some of
the possible NE. Three symmetric NE are

1.$\qquad \stackrel{\star }{p}=\stackrel{\star }{q}=0$ when $3\left|
b\right| ^{2}\leq 1$

2.$\qquad \stackrel{\star }{p}=\stackrel{\star }{q}=1$ when $3\left|
b\right| ^{2}\geq 2$

3.$\qquad \stackrel{\star }{p}=\stackrel{\star }{q}=3\left| b\right| ^{2}-1$
when $1<3\left| b\right| ^{2}<2\qquad \qquad (14)$

The first two NE are independent of the parameters $a$ and $b$ of the
initial state. However, the third NE depends on these. We now ask which of
these NE can be ESS's assuming that a particular NE exists with reference to
a particular set of initial states $\left| \psi _{in}\right\rangle $ for
which it can be found. The payoff to a player using $\stackrel{\curlywedge }{%
I}$ with probability $p$ when the opponent uses $\stackrel{\curlywedge }{I}$
with probability $q$ is

$P(p,q)=3\{pq\left| a\right| ^{2}+(1-p)(1-q)\left| b\right|
^{2}\}+5\{p(1-q)\left| b\right| ^{2}+q(1-p)\left| a\right| ^{2}\}$

$\qquad \qquad +\{pq\left| b\right| ^{2}+(1-p)(1-q)\left| a\right|
^{2}\}\qquad \qquad (15)$

For the first case $\stackrel{\star }{p}=\stackrel{\star }{q}=0$. The payoff 
$P(0,0)>P(p,0)$ when $3\left| b\right| ^{2}<1$ and $P(0,0)=P(p,0)$ imply $%
3\left| b\right| ^{2}=1$. Also $P(q,q)=-q^{2}+\frac{5}{3}(q+1)$ and $P(0,q)=%
\frac{5}{3}(q+1)$. Now $P(0,q)>P(q,q)$ when $q\neq 0.$Therefore, $\stackrel{%
\star }{p}=\stackrel{\star }{q}=0$ is an ESS when $3\left| b\right| ^{2}\leq
1$.

Consider $\stackrel{\star }{p}=\stackrel{\star }{q}=1$ now. $P(1,1)>P(p,1)$
means $3\left| b\right| ^{2}>2$ if $p\neq 1$. And $P(1,1)=P(p,1)$ means for $%
\ p\neq 1$ we have $3\left| b\right| ^{2}=2$. In such case $P(q,q)=-q^{2}+%
\frac{1}{3}(q+7)$ and $P(1,q)=\frac{5}{3}(2-q)$. Now $P(1,q)>P(q,q)$ because 
$(1-q)^{2}>0$ for $q\neq 1$. Therefore $\stackrel{\star }{p}=\stackrel{\star 
}{q}=1$ is an ESS when $3\left| b\right| ^{2}\geq 2$.

The third case $\stackrel{\star }{p}=\stackrel{\star }{q}=3\left| b\right|
^{2}-1$. Here $P(3\left| b\right| ^{2}-1,3\left| b\right| ^{2}-1)=-36\left|
b\right| ^{6}+36\left| b\right| ^{4}-5\left| b\right| ^{2}+6$. Also we find $%
P(p,3\left| b\right| ^{2}-1)=-21\left| b\right| ^{4}+21\left| b\right|
^{2}-3 $. Therefore, the condition $P(3\left| b\right| ^{2}-1,3\left|
b\right| ^{2}-1)>P(p,3\left| b\right| ^{2}-1)$ holds and $\stackrel{\star }{p%
}=\stackrel{\star }{q}=3\left| b\right| ^{2}-1$ is an ESS too for $1<3\left|
b\right| ^{2}<2$. All three possible symmetric NE definable for different
ranges of $\left| b\right| ^{2}$ turn out ESS's. Each of the three sets of
initial states $\left| \psi _{in}\right\rangle $\ give a unique NE that is
an ESS too. Switching from one to the other sets of initial states also
changes the NE and ESS accordingly. A question rises here: is it possible
that a particular NE switches over between `ESS' and `not ESS' when the
initial state changes between certain possible choices?. The transition
between classical and quantum game is also controlled by a change in the
initial state. For example classical payoffs can be obtained when the
initial state is unentangled. It implies that it may be possible to switch
over between `ESS' and `not ESS' by a change between `classical' and
`quantum' forms of a game i.e. when the initial state is unentangled and
entangled respectively. This possibility makes ESS interesting for the
quantum game theory as well. Because PD does not allow such a possibility we
now investigate asymmetric games to look for an answer for our question.

\section{Asymmetric case}

The players are anonymous in a symmetric bimatrix game. Such a game is
written as $G=(M,M^{T})$ where $M$ is a square matrix and $M^{T}$ is its
transpose. An ESS for an asymmetric bimatrix game i.e. $G=(M,N)$ when $N\neq
M^{T}$ is defined as a strict NE [11].A strategy pair $(\stackrel{\star }{x},%
\stackrel{\star }{y})\in S$, is an Evolutionarily Stable Strategy Pair of
the asymmetric bimatrix game $G=(M,N)$ if it satisfies the NE conditions
with strict inequality i.e.

1.$\qquad P_{A}(\stackrel{\star }{x},\stackrel{\star }{y})>P_{A}(x,\stackrel{%
\star }{y})$ for all $x\neq \stackrel{\star }{x}$

2.$\qquad P_{B}(\stackrel{\star }{x},\stackrel{\star }{y})>P_{B}(\stackrel{%
\star }{x},y)$ for all $y\neq \stackrel{\star }{y}\qquad \qquad (16)$

The game of Battle of Sexes has the following matrix

$\left( 
\begin{array}{cc}
(\alpha ,\beta ) & (\gamma ,\gamma ) \\ 
(\gamma ,\gamma ) & (\beta ,\alpha )
\end{array}
\right) \qquad \qquad $where $\alpha >\beta >\gamma \qquad \qquad (17)$

is an asymmetric bimatrix game with three classical NE [13]

1.$\qquad \stackrel{\star }{p_{1}}=\stackrel{\star }{q_{1}}=0$

2.$\qquad \stackrel{\star }{p_{2}}=\stackrel{\star }{q_{2}}=1$

3.\qquad $\stackrel{\star }{p_{3}}=\frac{\alpha -\gamma }{\alpha +\beta
-2\gamma }\qquad \stackrel{\star }{q_{3}}=\frac{\beta -\gamma }{\alpha
+\beta -2\gamma }\qquad \qquad (18)$

Here (1) and (2) are ESS's as well but (3) is not because it is not a strict
NE. The asymmetric quantum game played via the entangled state $\left| \psi
_{in}\right\rangle =a\left| OO\right\rangle +b\left| TT\right\rangle $,where 
$O$ and $T$ denote `Opera' and `Television' respectively, has following
three NE [13].

1.\qquad $\stackrel{\star }{p_{1}}=\stackrel{\star }{q_{1}}=1$

2.\qquad $\stackrel{\star }{p_{2}}=\stackrel{\star }{q_{2}}=0$

3.\qquad $\stackrel{\star }{p_{3}}=\frac{(\alpha -\gamma )\left| a\right|
^{2}+(\beta -\gamma )\left| b\right| ^{2}}{\alpha +\beta -2\gamma }\qquad 
\stackrel{\star }{q_{3}}=\frac{(\alpha -\gamma )\left| b\right| ^{2}+(\beta
-\gamma )\left| a\right| ^{2}}{\alpha +\beta -2\gamma }\qquad \qquad (19)$

Similar to the classical case, (1) and (2) are ESS's while (3) is not. These
two ESS's do not depend on the parameters $a$ and $b$ of the initial
state,however, the third NE does so. We show later that for other games an
ESS may also be dependent on the parameters $a$ and $b$. Interestingly
playing the Battle of Sexes game via another entangled state $\left| \psi
_{in}\right\rangle =a\left| OT\right\rangle +b\left| TO\right\rangle $
changes the scene. The payoffs to Alice and Bob now are:

$P_{A}(p,q)=$

$p\left\{ -q(\alpha +\beta -2\gamma )+\alpha \left| a\right| ^{2}+\beta
\left| b\right| ^{2}-\gamma \right\} +q\left\{ \alpha \left| b\right|
^{2}+\beta \left| a\right| ^{2}-\gamma \right\} +\gamma $

$P_{B}(p,q)=$

$q\left\{ -p(\alpha +\beta -2\gamma )+\beta \left| a\right| ^{2}+\alpha
\left| b\right| ^{2}-\gamma \right\} +p\left\{ \beta \left| b\right|
^{2}+\alpha \left| a\right| ^{2}-\gamma \right\} +\gamma \qquad \qquad (20)$

and there is only one NE that is not an ESS i.e.

$\stackrel{\star }{p}=\frac{\beta \left| a\right| ^{2}+\alpha \left|
b\right| ^{2}-\gamma }{\alpha +\beta -\gamma }\qquad \stackrel{\star }{q_{3}}%
=\frac{\alpha \left| a\right| ^{2}+\beta \left| b\right| ^{2}-\gamma }{%
\alpha +\beta -\gamma }\qquad \qquad (21)$

and Battle of Sexes playing via the state $\left| \psi _{in}\right\rangle
=a\left| OT\right\rangle +b\left| TO\right\rangle $ gives no ESS at all.

An essential requirement on a quantum version of a game is that the
corresponding classical game must be its subset. Suppose for a quantum game
corresponding to an asymmetric bimatrix classical game a particular strategy
pair $(\stackrel{\star }{x},\stackrel{\star }{y})\in S$ is an ESS
independent of an initial state $\left| \psi _{in}\right\rangle $ in its
possible choices i.e. $(\stackrel{\star }{x},\stackrel{\star }{y})$ is an
ESS for all $a$ and $b$.Classical game being a subset of the quantum game
the strategy pair $(\stackrel{\star }{x},\stackrel{\star }{y})$ must be an
ESS in the classical game as well. However, a strategy pair $(\stackrel{%
\star }{x},\stackrel{\star }{y})$ being an ESS in the classical game may not
remain an ESS in quantum version. The quantization of an asymmetric
classical game can make disappear the classical ESS's but cannot make appear
new ESS's, provided an ESS in quantum version remains so for every possible
choice of $a$ and $b$. However, when an ESS is defined as a strict NE
existing only for a set of initial states for which that NE exists the
statement that quantization can only make disappear classically available
ESS's may not remain valid. In such a case quantization may make appear new
ESS's definable for certain ranges of the parameters $a$ and $b$.To find
games with the property that ` a particular NE switches over between `ESS'
and `not ESS' when the initial state changes between its possible choices'
we write down the payoffs to Alice and Bob playing an asymmetric quantum
game via the method [13] of probabilistic choice of the operator $\stackrel{%
\curlywedge }{I}$ by the players. For the matrix

$\left( 
\begin{array}{cc}
(\alpha _{1},\alpha _{2}) & (\beta _{1},\beta _{2}) \\ 
(\gamma _{1},\gamma _{2}) & (\sigma _{1},\sigma _{2})
\end{array}
\right) \qquad \qquad (22)$

with the condition $\left( 
\begin{array}{cc}
\alpha _{1} & \beta _{1} \\ 
\gamma _{1} & \sigma _{1}
\end{array}
\right) \neq \left( 
\begin{array}{cc}
\alpha _{2} & \beta _{2} \\ 
\gamma _{2} & \sigma _{2}
\end{array}
\right) ^{T}$ the payoffs for the initial state $\left| \psi
_{in}\right\rangle =a\left| OO\right\rangle +b\left| TT\right\rangle $ are
as follows, given that $O$ and $T$ are again the two pure classical
strategies that no more represent `Opera' and `Television' only.

$P_{A}(p,q)=\alpha _{1}\left\{ pq\left| a\right| ^{2}+(1-p)(1-q)\left|
b\right| ^{2}\right\} +\beta _{1}\left\{ p(1-q)\left| a\right|
^{2}+q(1-p)\left| b\right| ^{2}\right\} $

$+\gamma _{1}\left\{ p(1-q)\left| b\right| ^{2}+q(1-p)\left| a\right|
^{2}\right\} +\sigma _{1}\left\{ pq\left| b\right| ^{2}+(1-p)(1-q)\left|
a\right| ^{2}\right\} $

$P_{B}(p,q)=\alpha _{2}\left\{ pq\left| a\right| ^{2}+(1-p)(1-q)\left|
b\right| ^{2}\right\} +\beta _{2}\left\{ p(1-q)\left| a\right|
^{2}+q(1-p)\left| b\right| ^{2}\right\} $

$+\gamma _{2}\left\{ p(1-q)\left| b\right| ^{2}+q(1-p)\left| a\right|
^{2}\right\} +\sigma _{2}\left\{ pq\left| b\right| ^{2}+(1-p)(1-q)\left|
a\right| ^{2}\right\} \qquad \qquad (23)$

The NE conditions are then

$P_{A}(\stackrel{\star }{p},\stackrel{\star }{q})-P_{A}(p,\stackrel{\star }{q%
})=$

$(\stackrel{\star }{p}-p)\left[ \left| a\right| ^{2}(\beta _{1}-\sigma
_{1})+\left| b\right| ^{2}(\gamma _{1}-\alpha _{1})-\stackrel{\star }{q}%
\left\{ (\beta _{1}-\sigma _{1})+(\gamma _{1}-\alpha _{1})\right\} \right]
\geq 0$

$P_{B}(\stackrel{\star }{p},\stackrel{\star }{q})-P_{B}(\stackrel{\star }{p}%
,q)=$

$(\stackrel{\star }{q}-q)\left[ \left| a\right| ^{2}(\gamma _{2}-\sigma
_{2})+\left| b\right| ^{2}(\beta _{2}-\alpha _{2})-\stackrel{\star }{p}%
\left\{ (\gamma _{2}-\sigma _{2})+(\beta _{2}-\alpha _{2})\right\} \right]
\geq 0\qquad \qquad (24)$

Let now $\stackrel{\star }{p}=\stackrel{\star }{q}=0$ be a NE i.e.

$P_{A}(0,0)-P_{A}(p,0)=-p\left[ (\beta _{1}-\sigma _{1})+\left| b\right|
^{2}\left\{ (\gamma _{1}-\alpha _{1})-(\beta _{1}-\sigma _{1})\right\} %
\right] \geq 0$

$P_{B}(0,0)-P_{B}(0,q)=-q\left[ (\gamma _{2}-\sigma _{2})+\left| b\right|
^{2}\left\{ (\beta _{2}-\alpha _{2})-(\gamma _{2}-\sigma _{2})\right\} %
\right] \geq 0\qquad \qquad (25)$

When the strategy pair $(0,0)$ is an ESS in the classical game $(\left|
b\right| ^{2}=0)$ we should have

$P_{A}(0,0)-P_{A}(p,0)=-p(\beta _{1}-\sigma _{1})>0$ for all $p\neq 0$ and

$P_{B}(0,0)-P_{B}(0,q)=-q(\gamma _{2}-\sigma _{2})>0$ for all $q\neq 0\qquad
\qquad (26)$

It implies $(\beta _{1}-\sigma _{1})<0$ and $(\gamma _{2}-\sigma _{2})<0$.

For the pair $(0,0)$ to be `not ESS' for some $\left| b\right| ^{2}\neq 0$
let take $\gamma _{1}=\alpha _{1\text{ }}$and $\beta _{2}=\alpha _{2}$ we
have then

$P_{A}(0,0)-P_{A}(p,0)=-p(\beta _{1}-\sigma _{1})\left\{ 1-\left| b\right|
^{2}\right\} $

$P_{B}(0,0)-P_{B}(0,q)=-q(\gamma _{2}-\sigma _{2})\left\{ 1-\left| b\right|
^{2}\right\} \qquad \qquad (27)$

And the pair $(0,0)$ doesn't remain an ESS when $\left| b\right| ^{2}=1$. A
game with these properties is given by the matrix

$\left( 
\begin{array}{cc}
(1,1) & (1,2) \\ 
(2,1) & (3,2)
\end{array}
\right) \qquad \qquad (28)$

For this game the pair $(0,0)$ is an ESS when $\left| b\right| ^{2}=0$
(classical game) but it is not when for example $\left| b\right| ^{2}=\frac{1%
}{2}$, though it remains a NE in both the cases. Therefore, a NE can be
switched between ESS and `not ESS' by adjusting the parameters $a$ and $b$.
An ESS may also appear when unentangled strategies become entangled opposite
to the previous case. An example of a game for which it happens is

$\left( 
\begin{array}{cc}
(2,1) & (1,0) \\ 
(1,0) & (1,0)
\end{array}
\right) \qquad \qquad (29)$

Playing this game again via $\left| \psi _{in}\right\rangle =a\left|
OO\right\rangle +b\left| TT\right\rangle $ gives following payoff
differences for the strategy pair $(0,0)$ for Alice and Bob respectively

$P_{A}(0,0)-P_{A}(p,0)=p\left| b\right| ^{2}$ \ \ and \ $\
P_{B}(0,0)-P_{B}(0,q)=q\left| b\right| ^{2}\qquad \qquad (30)$

Therefore $(29)$ is an example of a game for which $(0,0)$ is not an ESS
when initial state in unentangled but $(0,0)$ is an ESS for entangled
initial states i.e. $0<\left| b\right| ^{2}<1$.

\section{Concluding Remarks}

We have shown that in a population engaged in symmetric bimatrix classical
game of Prisoner's Dilemma an invasion of classical ESS is possible rather
easily by the mutants exploiting two-parameter set of quantum strategies.
However, the mutants cannot invade when they are deprived of using
entanglement or when entanglement doesn't remain an advantage. For an
asymmetric quantum game between two players we have shown that a strategy
pair can be made an ESS for either classical (using unentangled $\left| \psi
_{in}\right\rangle $) or quantum (using entangled $\left| \psi
_{in}\right\rangle $) version of the game even when the strategy pair
remains a Nash equilibrium in both the versions. It shows that in certain
types of games entanglement can be used to make appear or disappear ESS's
while retaining corresponding Nash equilibria.

The notion of an ESS in multiplayer classical games have been used in
classical game theory. Recently S.C.Benjamin [15] have shown that coherent
equilibria of mostly cooperative nature can exist in multiplayer quantum
games. We think that ESS can be useful refinement concept in multiplayer
quantum games having multiple NE and entanglement can also be related to
ESS's in these games as well.

\section{Acknowledgments}

One of us (A.I) is grateful to Pakistan Institute of Lasers and Optics,
Islamabad, for support. We are obliged to R. Naqvi for improving the text of
the paper.

\section{References}

[1] Maynard Smith, J. and Price, G.R. (1973) The logic of animal conflict.
Nature, 246, 15-18

[2] Maynard Smith, J. (1982) Evolution and the theory of games. CUP.

[3] Grafen, A (1990) Biological signals as handicaps. J. Theor. Bio. 144,
517-546

[4] D.Meyer, Phys. Rev. Lett. 82, 1052 (1999) and quant-ph/0004092

[5] J.Eisert, M. Wilkens, and M. Lewenstein, Phys. Rev. Lett. 83, 3077
(1999) and quant-ph/0004076

[6] R. Dawkins, The Selfish Gene (Oxford University Press, Oxford, 1976).

[7] R.B. Myerson, Game Theory: An Analysis of Conflict (MIT Press,
Cambridge, MA, 1991).

[8] Game Theory, A report by K. Prestwich, Department of Biology, College of
the Holy Cross, Worcester, MA, USA 01610. 1999.

[9] Evolution in knockout conflicts. A report by M.Broom, Center for
Statistics and Stochastic Modeling, School of Mathematical Sciences,
University of Sussex, U.K. September 17, 1997.

[10] States, Effects, and Operations: Fundamental Notions of Quantum Theory,
by K.Kraus. Lecture Notes in Physics, Vol. 190. (Springer-Verlag, Berlin,
1983).

[11] Evolutionary game theory and the Modelling of Economic Behaviour, by
Gerard van der Laan and Xander Tieman. November 6, 1996. Research Program
''Competition and Cooperation'' of the Faculty of Economics and
Econometrics, Free University, Amsterdam.

[12] Reply to ``Comment on: A Quantum Approach to Static Games of Complete
Information'' quant-ph/0009103

[13] A Quantum Approach to Static Games of Complete Information. Phys. Lett,
272, 291 (2000);

[14] Comment on: ``A quantum approach to static games of complete
information''

quant-ph/0008127

[15] Multiplayer Quantum Games. quant-ph/0007038

\end{document}